\documentclass[modern]{aastex6}
\usepackage[T1]{fontenc}
\usepackage[latin9]{inputenc}
\usepackage{xcolor}
\usepackage{textcomp}
\usepackage{graphicx}
\usepackage{epstopdf}

\makeatletter

\shorttitle{}
\shortauthors{}

\usepackage[american]{babel}
\usepackage{amsmath}

\makeatother

\begin{document}

\title{TYCHO -- Realistically Simulating Exoplanets within Stellar Clusters
\\
I: Improving the Monte Carlo Approach}

\author{Joseph P. Glaser}
\affiliation{Department of Physics, Drexel University, Disque Hall 816, 32 S. 32nd Street,Philadelphia, PA 19104, U.S.A}
\email{joseph.p.glaser@drexel.edu}

\author{Stephen L.W. McMillan}
\affiliation{Department of Physics, Drexel University, Disque Hall 816, 32 S. 32nd Street,Philadelphia, PA 19104, U.S.A}

\author{Aaron M. Geller}
\affiliation{Center for Interdisciplinary Exploration and Research in Astrophysics (CIERA) and Department of Physics and Astronomy, Northwestern University, 1800 Sherman Ave., Evanston, IL 60201, USA} 
\affiliation{Adler Planetarium, Department of Astronomy, 1300 S. Lake Shore Drive, Chicago, IL 60605, USA}

\author{Jonathan D. Thornton}
\affiliation{Department of Physics, Drexel University, Disque Hall 816, 32 S. 32nd Street,Philadelphia, PA 19104, U.S.A}

\author{Mark R. Giovinazzi}
\affiliation{Department of Physics and Astronomy, University of Pennsylvania, 209 S 33rd Street, Philadelphia, PA 19104, USA}

\begin{abstract}
To fully understand the diverse population of exoplanets, we must
study their early lives within open clusters, the birthplace
of most stars with masses $>0.5M_\odot$ (including those currently in the field). Indeed, when we observe planets within clustered
environments, we notice highly eccentric and odd systems that
suggest the importance of dynamical pathways created by interactions
with additional bodies (as in the case of HD 285507b). However,
it has proven difficult to investigate these effects, as many current
numerical solvers for the multi-scale $N$-body problem are simplified
and limited in scope. To remedy this, we aim to create a physically complete computational
solution to explore the role of stellar close encounters and interplanetary
interactions in producing the observed exoplanet populations for both
open cluster stars and field stars. We present a new code, \textsc{Tycho},
which employs a variety of different computational techniques, including
multiple $N$-body integration methods, close encounter handling,
modified Monte Carlo scattering experiments, and a variety of empirically
informed initial conditions. We discuss the methodology in
detail, and its implementation within the AMUSE software framework. Approximately 1\% of our systems are promptly disrupted by star-star encounters contributing to the rogue planets occurrence rate. Additionally, we find that close encounters which perturb long-period planets lead to 38.3\% of solar-system-like planetary systems becoming long-term unstable.
\end{abstract}


\section{Introduction}
The rate of exoplanet discoveries has exploded recently, fueled by
new telescopes, technological advances, and increased interest
among both astronomers and the general public. Discoveries of exoplanets
in the Galactic field have dramatically outpaced discoveries in star
clusters. This may be due to the limited number of stars observed
by exoplanet surveys in star clusters, leading to an observational
bias \citep{VanSaders2011}. However, over the past several years, 
there have been a number of detections of planets in open star clusters, many with
orbital parameters unlike what we see around field stars.
As of this work's publication, there have been 34 planets observed
within 10 Galactic open clusters. In Table \ref{tab:exoplanet_cat} we 
catalog these planets and their parent cluster's relevant parameters.
Additionally, Table \ref{tab:cluster_cat}
presents the measured parameters for the open clusters mentioned.
From hot Jovians to eccentric Neptunes to volatile Super Earths, these
exoplanets display the vast diversity of the planetary population.
With new telescopes coming online and continued interest in exoplanet
science, new discoveries of exoplanets around open cluster stars are
inevitable.

\floattable
\begin{deluxetable}{lccccccc}
    \tabletypesize{\footnotesize{}}
    \tablecaption{\label{tab:exoplanet_cat}Parameters of Confirmed Exoplanets Located Within Galactic Open Clusters}
    \tablehead{
    \colhead{Designation} & \colhead{$M_{\star}$ $\left(M_{\odot}\right)$} & \colhead{$M_{p}$ $\left(M_{J}\right)$} & \colhead{$R_{p}$ ($R_{J}$)} & \colhead{$P$ (Days)} & \colhead{$a$ (AU)} & \colhead{$e$} & \colhead{Ref.}
    }
    \startdata
    	\multicolumn{8}{l}{\emph{Praesepe}} \\
    Pr 0201 b  & $1.24\pm0.039$ & $1.534_{-0.043}^{+0.038}$ & - & $4.4264\pm0.007$ & - & $0.079\pm0.078$ & A \\
    Pr 0211 b & $0.935\pm0.013$ & $1.88\pm0.02$ & - & $4.4264\pm3.0E-5$ & $0.03184\pm1.5E-4$ & $0.017\pm0.01$ & A \\
    Pr 0211 c & " & $7.79\pm0.33$ & - & $4850_{-1750}^{+4560}$ & $5.5_{-1.4}^{+3.0}$ & $0.71\pm0.11$ & B, C \\
    K2-95 b & $0.44\pm0.01$ & $1.67_{-1.66}^{+0.0}$ & $0.33\pm0.018$ & $10.1340\pm0.0011$ & $0.0653_{-0.0045}^{+0.0187}$ & $0.16_{-0.11}^{+0.19}$ & D, E \\
    K2-100 b & $1.18\pm0.09$ & $11.81_{-5.34}^{+9.3} M_{\oplus}$ & $0.31\pm0.018$ & $1.67391\pm0.0011$ & $0.0296\pm0.0002$ & $0.24_{-0.12}^{+0.19}$ & D, F \\
    K2-101 b & $0.80\pm0.06$ & - & $0.18\pm0.01$ & $14.6773\pm0.0008$ & - & $0.10_{-0.08}^{+0.18}$ & D, F \\
    K2-102 b & $0.77\pm0.06$ & - & $0.12\pm0.01$ & $9.91562\pm0.0012$ & - & $0.10_{-0.07}^{+0.16}$ & D, F \\
    K2-103 b & $0.61\pm0.02$ & - & $0.196_{-0.009}^{+0.018}$ & $21.1696\pm0.0017$ & - & $0.18_{-0.15}^{+0.27}$ & D, F \\
    K2-104 b & $0.51\pm0.02$ & - & $0.170_{-0.018}^{+0.009}$ & $1.97424\pm0.0001$ & - & $0.18_{-0.14}^{+0.29}$ & D, F \\
    EPIC 211901114 b & $0.46\pm0.02$ & $<5.0$ & $0.9_{-0.4}^{+0.5}$ & $1.64893\pm0.0001$ & - & - & D, F \\
    Pr 0157 b & $\sim0.9$ & $9.49\pm0.66$ & - & $1234\pm11$  & - & $0.577\pm0.023$ & C \\
    \hline
	    \multicolumn{8}{l}{\emph{Hyades}} \\
    $\epsilon$ Tau b & $2.7\pm0.01$ & $7.6\pm0.2$ & - & $549.9\pm5.3$  & $1.93\pm0.03$ & $0.151\pm0.023$ & G \\
    HD 285507 b & $0.734\pm0.034$ & $9.17\pm0.033$ & - & $6.0881\pm0.0018$  & $0.0729\pm0.003$ & $0.086\pm0.019$ & H \\
    K2-25 / aV 50 b & $0.261\pm0.021$ & $<1.15$ & $0.306_{-0.028}^{+0.085}$  & $3.4845_{-3.7E-5}^{+3.1E-5}$ & - & $0.27_{-0.21}^{+0.16}$ & I \\
    K2-136 b & $0.74\pm0.02$ & - & $0.0883_{-0.0036}^{+0.0054}$ & $7.975292_{-0.00077}^{+0.00083}$ & - & $0.10_{-0.07}^{+0.19}$ & J \\
    K2-136 c & " & - & $0.260_{-0.009}^{+0.010}$ & $17.30714_{-0.00028}^{+0.00025}$ & - & $0.13_{-0.11}^{+0.27}$ & J \\
    K2-136 d & " & - & $0.129_{-0.007}^{+0.010}$ & $25.57507_{-0.00236}^{+0.00242}$ & - & $0.14_{-0.09}^{+0.13}$ & J \\
    HD 283869 b & $0.74\pm0.03$ & - & $0.196\pm0.13$ & $106.0_{-25}^{+74}$ & - & - & K \\
    \hline
        \multicolumn{8}{l}{\emph{M67}} \\
    NGC 2682 YBP 1194 b & $1.01\pm0.02$ & $0.32\pm0.03$ & - & $6.959\pm0.001$ & - & $0.28\pm0.08$ & L \\
    NGC 2682 YBP 1514 b & $0.96\pm0.01$ & $0.40\pm0.38$ & - & $4.087\pm0.001$ & - & $0.28\pm0.09$ & L \\
    NGC 2682 SAND 364 b & $1.35\pm0.05$ & $1.57\pm0.11$ & - & $120.951\pm0.453$ & - & $0.35\pm0.10$ & L \\
    NGC 2682 SAND 978 b & $1.37\pm0.02$ & $2.18\pm0.17$ & - & $511.21\pm2.04$ & - & $0.16\pm0.07$ & L \\
    NGC 2682 YBP 401 b & $1.14\pm0.02$ & $0.42\pm0.05$ & - & $4.087\pm0.007$ & - & $0.16\pm0.08$ & M \\
    \hline
        \multicolumn{8}{l}{\emph{Coma Berenices}} \\
    CB0036 b & $\sim1.2$ & $2.49\pm0.10$ & - & $43.831_{-0.047}^{+0.039}$ & - & $0.239_{-0.034}^{+0.027}$ & C \\
    CB0036 c & " & $2.7_{-0.4}^{+1.5}$ & - & $449_{-12}^{+13}$ & - & $0.69_{-0.04}^{+0.20}$ & C \\
    \hline
	    \multicolumn{8}{l}{\emph{NGC 6811}} \\
    Kepler-66 b & $1.038\pm0.044$ & - & $0.250\pm0.014$ & $17.816\pm7.5E-5$ & $0.1352\pm0.0017$ & - & N \\
    Kepler-67 b & $0.865\pm0.034$ & - & $0.260\pm0.014$ & $15.726\pm1.1E-4$ & $0.1171\pm0.0015$ & - & N \\
    \hline
	    \multicolumn{8}{l}{\emph{Upper Scorpius}} \\
    K2-33 b & $0.54_{-0.007}^{+0.011}$ & $<3.6$ & $0.451\pm0.033$ & $5.42513\pm0.0003$ & $0.0409\pm0.023$ & - & O \\
    1RXS J1609\-2105 b & $0.85_{-0.10}^{+0.20}$ & $14.0_{-3.0}^{+2.0}$  & - & - & $<330$ & - & P \\
    GSC 6214-210 b & $0.83\pm0.02$ & $14.5\pm2.0$ & - & - & $<320$ & - & Q \\
    \hline
	    \multicolumn{8}{l}{\emph{Ruprecht}}\\
    EPIC 219388192 b & $0.99\pm0.05$ & $36.5\pm0.009$ & $0.937\pm0.042$ & $5.2926\pm2.6E-5$ & $0.0593\pm0.0029$ & $0.193\pm0.002$ & R \\
    \hline
   	    \multicolumn{8}{l}{\emph{NGC 4349}} \\
    127 b & $3.81\pm0.23$ & $\sim24.097$ & - & $671.94\pm5.32$ & $\sim2.35$ & $0.046\pm0.033$ & S \\
    \hline
  	    \multicolumn{8}{l}{\emph{NGC 2423}} \\
    3 b & $2.26\pm0.07$ & $\sim9.621$ & - & $698.61\pm2.72$ & $\sim2.02$ & $0.088\pm0.041$ & S \\
    \hline
   	    \multicolumn{8}{l}{\emph{IC 4651}} \\
    9122 b & $2.06\pm0.09$ & $\sim7.202$ & - & $747.22\pm2.95$ & $\sim2.05$ & $0.150\pm0.068$ & T \\
    \enddata
    \tablerefs{A \citep{Quinn2012}; B \citep{Malavolta2016}; C \citep{Quinn2016}; D \citep{Obermeier2016}; E \citep{Mann2017a}; F \citep{Stefansson2018};
G \citep{Sato2007}; H \citep{Quinn2014}; I \citep{Mann2016, David2016}; J \citep{Mann2017b,Livingston2017}; K \citep{Vanderburg2018}; 
L \citep{Brucalassi2014}; M \citep{Brucalassi2016}; N \citep{Meibom2013}; O \citep{Mann2016a,David2016a}; P \citep{Lafreniere2008,Lafreniere2010}; 
Q \citep{Ireland2011,Pearce2018}; R \citep{Nowak2018}; S \citep{Lovis2007}; T \citep{Mena2018}}
    \tablecomments{The above columns are as follows: the host star's mass, $M_{\star}$; the planet's mass (or $m\sin{i}$), $M_{p}$; the planet's radius, $R_{p}$; the orbital period, $P$; the orbital semimajor axis, $a$; and the orbital eccentricity. Due to the limitations of different detection methods and the differences in stellar magnitude sensitivity in surveys, some planets currently do not have certain parameters measured. Predictions and upper bounds have been provided where possible.
This collection has been drawn from the following sources: the\textit{
Extrasolar Planets Encyclopaedia} (exoplanet.eu); the NASA Exoplanet
Archive; and the Open Exoplanet Catalogue \citep[www.openexoplanetcatalogue.com]{Rein2012}.}
\end{deluxetable}
\floattable
\begin{deluxetable}{lcccccc}
    \tablecaption{\label{tab:cluster_cat}Parameters of Open Clusters with Detected Exoplanets}
    \tablehead{
    \colhead{Designation} & \colhead{$M_{T}$ $\left(M_{\odot}\right)$} & \colhead{$t_{age}$ (Myr)} & \colhead{$D_{\odot}$ (kpc)} & \colhead{$R_{c}$ (pc)} & \colhead{$R_{T}$ (pc)} & \colhead{[Fe/H]}
    }
    \startdata
    Hyades & $\sim400$ & $625\pm50$ & $0.04634\pm0.0003$ & $\sim2.7$ & $\sim10.3$ & $+0.146\pm0.004$ \\
    Praesepe / NGC 2632 & $511\pm73$ & $830$ & $\sim0.187$ & $\sim3.5$ & $\sim12$ & $+0.156\pm0.004$ \\
    M67 / NGC 2682 & ${2100}^{610}_{550}$ & $3550$ & $0.841\pm0.048$ & $0.60\pm0.06$ & $4.13\pm0.43$ & $+0.050\pm 0.040$ \\
    Coma Berenices / Melotte 111 & $112\pm16$ & $690$ & $\sim0.087$ & $6.80\pm0.30$ & $<12.85$ & $+0.070\pm0.090$ \\
    NGC 6811 & $144\pm32$ & $710$ & $1.460\pm0.086$ & $0.36\pm0.04$ & $2.89\pm0.42$ & $-0.500\pm0.200$ \\
    Upper Scorpius / Sco OB2-2 & $\sim2060$ & $11\pm2$ & $\sim0.145$ & $\sim28$ & $\sim50$ & - \\
    Ruprecht 147 & $58.88\pm1.45$ & $2140$ & $0.295\pm0.015$ & $6.51\pm3.78$ & $6.76\pm0.97$ & $+0.070\pm0.030$ \\
    NGC 4349 & $1654\pm199$ & $710$ & $1.430\pm0.085$ & $0.64\pm0.09$ & $3.68\pm0.58$ & $+0.105\pm0.108$ \\
    NGC 2423 & $120\pm30$ & $1000$ & $0.756\pm0.044$ & $0.18\pm0.03$ & $1.23\pm0.20$ & $+0.068\pm0.103$ \\
    IC 4651 / Melotte 169 & $1443\pm277$ & $1410$ & $1.006\pm0.059$ & $0.70\pm0.11$ & $3.48\pm0.36$ & $-0.128\pm0.082$ \\
    \enddata
    \tablerefs{Hyades \citep{Perryman1997,Cummings2017};
Praesepe \citep{Adams2002,Kharchenko2013,Cummings2017}; M67 \citep{Bukowiecki2011,Kharchenko2013, Geller2015,Overbeek2016}; NGC
4349, NGC 2423, \& IC 4651 \citep{Bukowiecki2011,Kharchenko2013};
Coma Berenices \citep{Casewell2006,Kraus2007,Melnikov2012,Kharchenko2013,Joshi2016};
NGC 6811 \citep{Bukowiecki2011}; Upper Scorpius \citep{DeZeeuw1999,Preibisch2008,Pecaut2012,Donaldson2017};
and Ruprecht 147 \citep{Kharchenko2013,Curtis2013,Joshi2016}.}
    \tablecomments{The above columns are as follows: total cluster mass, $M_T$; estimated age from isochrone fits, $t_{age}$; distance from the Sun to the cluster's core, $D_{odot}$; the core radius, $R_c$; the tidal radius, $R_T$; and the metallicity of the cluster compared to the solar standard, [Fe/H]. The most recent measurements for each
parameter have been recorded in this table assuming the values agree with the guidelines laid out in \citet{Netopil2015}.}
\end{deluxetable}

Most stars with masses $>0.5M_{\odot}$ were likely
born in clustered environments; a substantial fraction of these
natal clusters have since dissolved
in the Galactic tidal field \citep{Lada2010,Zwart2010,Fujii2015}.
Therefore, many planetary systems currently observed in the
Galactic field were likely born in significantly denser environments.
Fully understanding the origins of the orbital properties of exoplanet
systems will require a detailed study of their dynamical histories
from their birth cluster onward. Modeling the dynamics of planetary
systems in star clusters is a complex problem, largely due to the
very different timescales involved, beginning with planetary orbits of order
days through stellar orbits throughout the cluster of order Myr. In this
paper we develop the modeling tools needed to accomplish this task.

Coupling planetary orbit evolution and cluster dynamics is a complex
multi-scale problem that cannot be efficiently solved using conventional
numerical methods. Previous studies have used direct $N$-body calculations
\citep{Spurzem2009}, Monte-Carlo schemes for modeling close encounters
\citep{Hao2013}, and simulating gravitational potential perturbations
on planetary systems from the cluster \citep{Cai2014}. Direct $N$-body
calculations can be computationally costly and may pose scaling
problems when applied to large systems \citep{Aarseth2003}. Monte
Carlo approaches to simulating the close encounter histories of a
system are fast, but may not adequately represent conditions in clusters with small-$N$, like the open clusters we are interested in here \citep{Fregeau2003,Giersz2013}. Linking the gravitational
potential of a cluster to separate simulations of the planetary systems
running in parallel, while highly accurate, can only be implemented
for short cluster lifetimes due to computational limitations. Therefore, a numerical method that allows for
the accurate determination of cluster dynamics and planetary orbital
evolution over long time periods (comparable to the dissolving time
of an open cluster) with low computational overhead is required.

In this paper we describe such an implemented methodology, called
\textsc{Tycho}, and provide proof-of-concept results. In Section
2 we discuss \textsc{Tycho}'s implementation within the AMUSE software
framework \citep{Zwart2017}.\footnote{The Python-based interface is a
community effort with its main development
team based at the Leiden Observatory. It is available freely via its
GitHub repository at \url{https://github.com/amusecode/amuse}.} In Section 3, we detail our chosen initial conditions, which draw
heavily from recent empirical studies. In Sections 4 and 5, we explore
some representative results of production runs illustrating \textsc{Tycho}'s
current capabilities.

\begin{figure}[!p]
\centering{}\includegraphics[height=0.9\textheight]{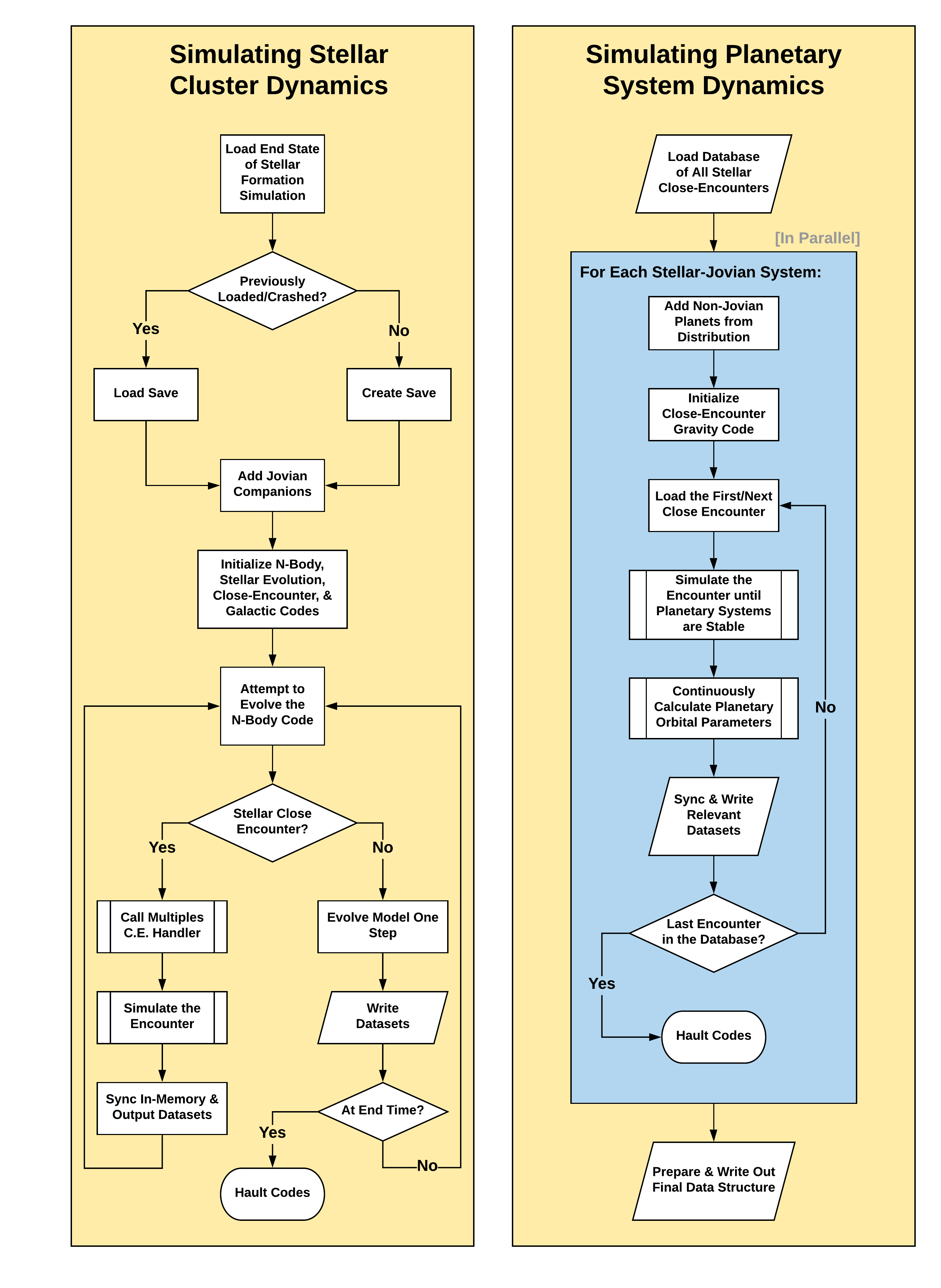}
\caption{\label{fig:flowchart}Flow chart of the general processes involved
in \textsc{Tycho} simulation pipeline. }
\end{figure}

\section{Methodology}

In this section, we describe in detail the methods used in
\textsc{Tycho}. In short, we record the close encounter histories
of stars via direct $N$-body simulations of clusters, which inform
a series of scattering experiments on multi-planet systems contained
within the cluster. Utilizing multiple independently-verified integrators
and packages, the result is a robust method for accurately studying
the change of planetary systems due to all dynamical effects present
in such environments. For a visual depiction of our methodology, we
refer the reader to Figure \ref{fig:flowchart}.

\subsection{Modeling the Cluster's $N$-body Dynamics
\label{subsec:Modeling-Dynamics}}

Cluster-scale gravitational dynamics is handled by \textsc{PH4}, an
MPI-parallel 4th-order Hermite direct $N$-body integrator with GPU
acceleration \citep{McMillan2011}. This AMUSE module is well tested
on $N$-body systems of all sizes, from binaries to clusters containing
hundreds of thousands of stars (see \citealt{Zwart2017}). Like most
$N$-body integrators in AMUSE, \textsc{PH4} does not include any
special treatment of close encounters between stars or binaries. Most
AMUSE modules, including PH4, include the possibility of softening
the potential, but this is not appropriate for the collisional dynamics
of interest here. Instead, \textsc{Tycho} uses the AMUSE \textsc{Multiples}
module to manage dynamical close encounters
\citep{PelupessyElterenVriesEtAl2013,Zwart2017}, as we now describe.

Traditional collisional $N$-body codes devote a substantial fraction
of the total code base to modeling the detailed internal motion of
multiple systems (where this term henceforth includes stellar binaries)
simultaneously with the large-scale motion of the rest of the system
\citep{PortegiesZwart1998,Aarseth2003}. The algorithms involved have
varying degrees of efficiency and accuracy, but all gradually relax
the accuracy of the perturbed binary integration until ``unperturbed''
binaries (where the external tidal acceleration is less than $\sim10^{-6}$
times any internal acceleration) are treated as isolated unperturbed
objects. The alternative approach, widely used in Monte Carlo schemes,
treats stable multiples as completely unperturbed until they interact
strongly with another object \citep{Fregeau2004,Tanikawa2009,Giersz2013}.

As discussed in more detail by \citet[Sec. 4.5]{Zwart2017}, the $N$-body
module (here \textsc{PH4}) manages a multiple's center of mass, but
is ignorant of its internal dynamics. The \textsc{Multiples} module
maintains a database of all stable multiples in the system.
Herein, we define that a stable multiple is a binary, a hierarchical triple deemed stable
by \citet{Mardling2007} criterion, or a higher-order multiple system
in which the Mardling criterion applied to the outermost orbit(s)
indicates approximate stability. Each single or multiple object
has an interaction radius $R_{int}$
assigned to it. For a single star of mass $M$, $R_{int}\sim GM/\sigma^{2}$,
where $\sigma$ is the cluster velocity dispersion. For a multiple,
$R_{int}$ is twice the outer semi-major axis. For
planets, $R_{int}$ is the Hill radius when bound to a star, and
$R_{int}\sim GM/\sigma^{2}$
when free-floating. When two particles in the $N$-body code approach
within the sum of their interaction radii, the $N$-body integrator
is advanced to and paused at that moment. The particles are then removed
from the $N$-body system, their internal structure is restored, and
the interaction is followed as an isolated few-body system using the
\textsc{SmallN} integrator, a robust shared-step time-symmetrized
code with analytic extensions \citep{Hut1995,McMillan1996,Zwart2017}.
Pure two-body encounters are advanced analytically using the AMUSE
\textsc{Kepler} module.

In order to inform the high-fidelity scattering experiments detailed
in Section \ref{subsec:Scattering-Experiments}, we 
store the encounter history of each star throughout the course of the
simulation. At the start of a close encounter, \textsc{Tycho} copies
the flattened hierarchical tree of the root particles involved 
into an AMUSE particle set. This particle set and all required initial
conditions are then stored in a chronological database of encounter
histories for each star in the simulation.
Once the $N$-body evolution ends, a complete
history of all stellar close encounters is available within the
database.

\begin{figure}[!p]
\centering{}\includegraphics[height=0.4\textheight]{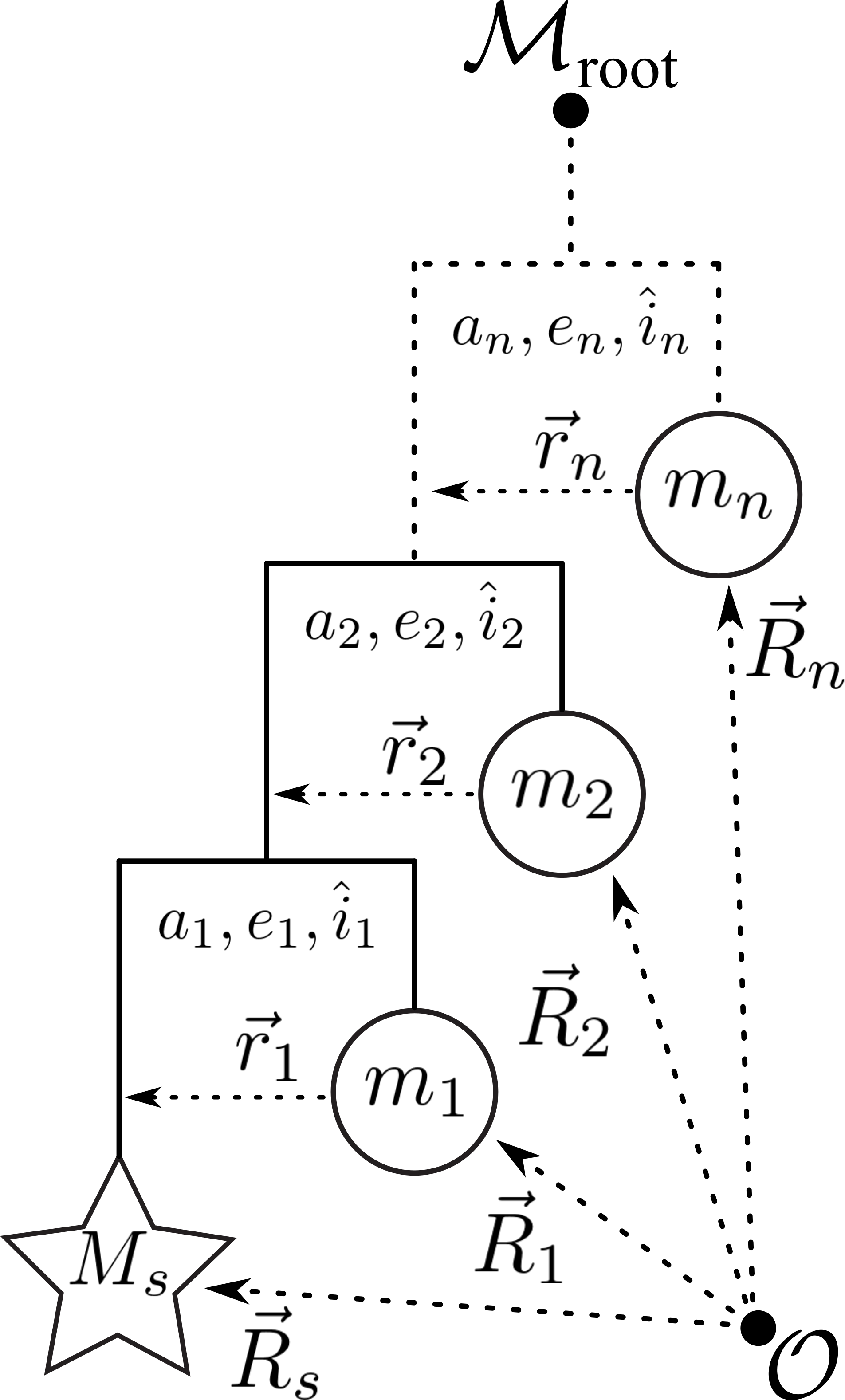}\caption{\label{fig:hierarchy}
A typical hierarchical system
in our simulations. \textsc{Multiples} handles such systems
by grouping nearest bound neighbors in a tree structure, storing
all particle information at each level. This process continues until
there are no further bound pairs. Not shown here is the fact that
the leaves of the tree can themselves be roots of their own trees.}
\end{figure}

The structure of the few-body system is monitored during the \textsc{SmallN}
integration and the integration is declared over when the system consists
of some number of mutually unbound, receding, stable (as just defined)
centers of mass. For stars with planetary systems, the ``over''
criterion is applied only to the stars. At that point, the internal
structure of the newly obtained multiples is stored for future use,
corrections are applied to partially account for the tidal field of
the rest of the $N$-body system, the centers of mass are added
to the $N$-body system, and the large-scale integration continues.
The internal structure of each center of mass particle is held static
until the next interaction occurs. Examples of this hierarchical structure
can be found in Figure \ref{fig:hierarchy}.

\subsection{Stellar Evolution}

For our models, all primary stellar masses are drawn from a \citet{Kroupa2001} mass function, as
discussed in detail in Section \ref{subsec:Stars-and-Primordial}. In addition, we
include primordial binaries and allow dynamical binaries to form
naturally, to ensure an accurate depiction of large-scale cluster
dynamics. \textsc{Tycho} uses the \textsc{SeBa} stellar evolution
module (with solar metallicity), with an additional
prescription for mass loss due to winds 
\citep{Portegies_SeBa,Tooen_Portegies_Zwart_Seba,PortegiesZwart2012}.

\begin{figure}
\centering{}\includegraphics[height=0.4\textheight]{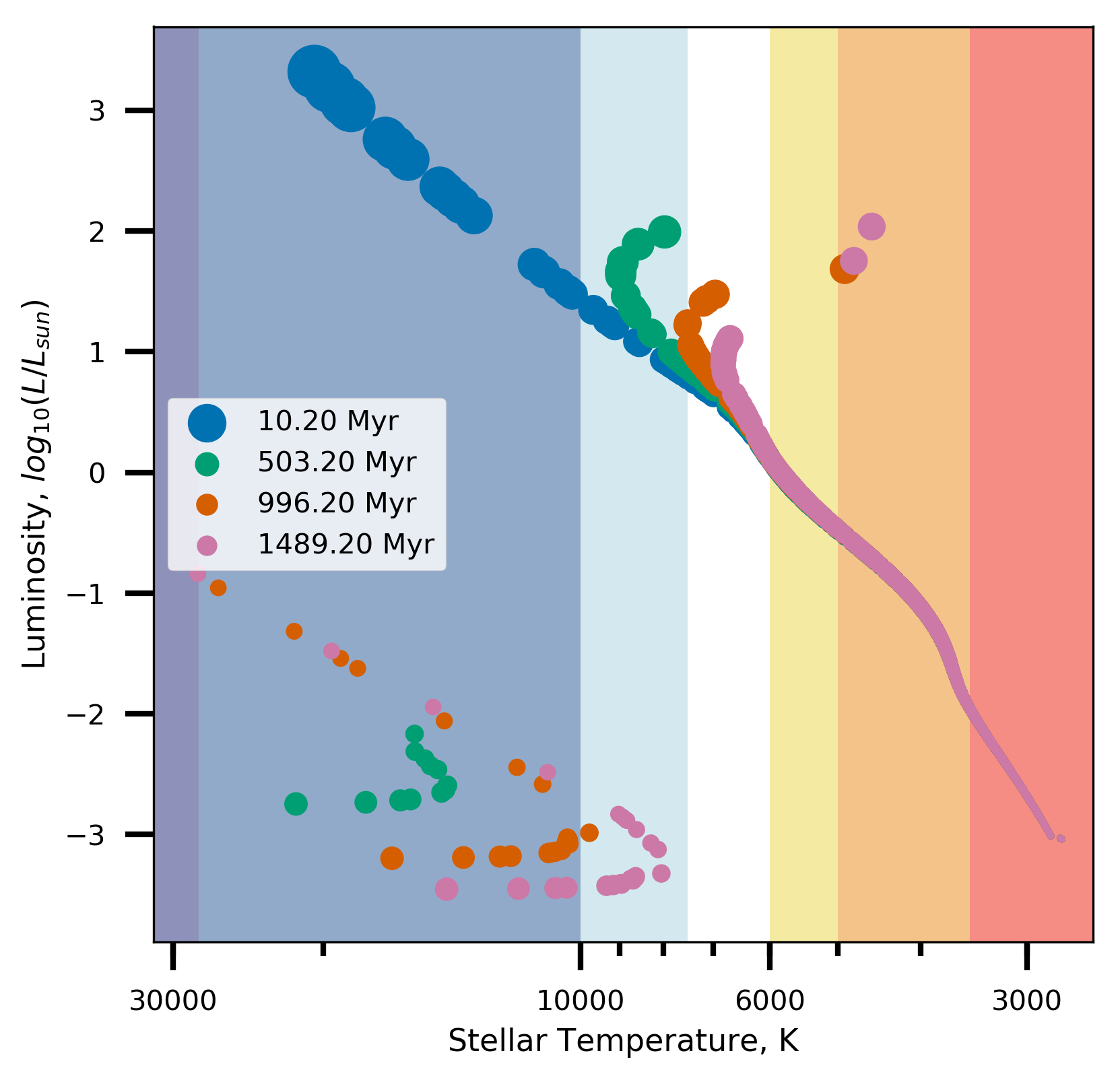}\caption{\label{fig:HR}Hertzsprung-Russell diagram
for a $\sim350 M_{\odot}$ ($\sim675$ stars) 
cluster at various epochs. Each
point represents an individual
star whether or not the star is part of a close hierarchical
system. By 1.5 Gyr, we have lost $\sim50 M_{\odot}$ to stellar evolution and supernovae.
}
\end{figure}

To account for multiple age groups in our stellar population, we implement
a procedure for asynchronous evolution within \textsc{SeBa}. This
methodology will become increasingly more important for accurate depictions
of dynamics as direct simulations of cluster formation from molecular
clouds become more widespread \citep[e.g.][]{Wall2018,Wall2019Formation}. In order to accomplish this, we bin our
stars by their age, which allows us to create separate instances of
\textsc{SeBa} for each grouping. These separate instances within the
AMUSE environment allow for the module to accurately evolve while
reducing the computational overhead to a manageable level. See Figure
\ref{fig:HR} for an example of a Hertzsprung-Russell
diagram drawn from a roughly 500 $M_{\odot}$ cluster.

\subsection{The Background Galactic Potential}

The evolution of a cluster within the tidal field of its host galaxy
is substantially different from one in isolation. The tidal field imposes a
limiting Jacobi radius on the cluster, at
which stars become unbound \citep{VonHoerner1957,King1962}.
As mass is lost from the cluster, the radius shrinks,
leading to a decreased cluster lifetime \citep{Baumgardt2003,Gieles2008}.
In order to accurately model the escape of stars from the cluster
into the Galactic field, we apply a background gravitational potential
which models key aspects of the Milky Way Galaxy. As described in detail in \citet{Bovy2015},
this static potential features a central bulge, a dusty disk and a
dark matter halo modeled after current Milky Way observations. To
couple the Galactic potential with the time-dependent evolution of
the cluster, we utilize the AMUSE framework's \textsc{Bridge} package,
a code-coupling algorithm based on the scheme described by \citet{Fujii2007}.
A cluster may be placed in any kind of orbit in this potential
and the relevant tidal forces will be fully taken into effect.

\subsection{Handling Planetary Dynamics on Different Scales
\label{subsec:Scattering-Experiments}}

Of the 4126 confirmed exoplanets to date,
1766 exist within 705 multi-planetary systems.\footnote{Statistics retrieved from the NASA Exoplanet
Archive on February 18th,
2020 drawing from the Kepler, K2, KELT, SuperWASP, TESS, and UKIRT surveys in addition to a number of ground-based observatories and related surveys.}
Accordingly, it is critical to include within \textsc{Tycho} the
ability to faithfully model multi-planet systems. In our model, we
record the parameters of all close encounters between stellar systems
(single or hierarchical) in the cluster and use these parameters to
later perform a series of scattering experiments, the procedure for which we will now describe.

Once the full-cluster $N$-body simulations are completed, we re-simulate our independent
encounters in parallel at higher temporal resolution with additional
planetary bodies drawn from our mock solar-system
model described in Section \ref{subsec:PlanetaryIC}. We again utilize the
specialized few-body
integrator \textsc{SmallN} for these scattering experiments. To
improve run-time efficiency, we advance the two stellar
systems involved in the encounter along their Keplerian orbit to
the point, $r_{12}$, where
the gravitational perturbation on the outermost orbit is of order
$10^{-3}$: 
\begin{equation}
r_{12}\sim10 a_{outer}\frac{M_{pert}}{M_{host}}.
\end{equation}
Here $a_{outer}$ is the semi-major axis of the outermost planet,
$M_{pert}$ is the total mass of the perturbing system, and $M_{host}$ is
the mass of the star hosting the planetary system. We then allow \textsc{SmallN}
to simulate the close encounter utilizing an adaptive internal time step,
storing particle position and velocity vectors every
1 year until the close encounter is declared over, as described in
Section \ref{subsec:Modeling-Dynamics}.
We then  evolve the system for an additional 100 years before committing and tabulating the planets' orbital parameters.

To ensure that our dynamical parameters are fully sampled, we simulate
each encounter in our database with the host star's planetary system
multiple times, with different orientations and mean anomalies. At the
start of each scattering experiment instance, we apply a random rotation
transformation which samples the unit sphere uniformly to the system
using the fast algorithm developed by \citet{Arvo1992}. The proof of
this method's uniform coverage is outlined by \citet{Shoemake1992}.
To develop a statistically significant population of encounters from
our $N$-body database, we repeat this procedure 100 times.

\subsection{Determining Long-Term Stability of Planetary Systems}
\label{subsec:AMD}
As we are interested in the evolution of our systems over their time spent within the parent cluster, we need a quantifiable system for analyzing not only the shifts in a given planet's orbits, but also the overall stability of the system they inhabit. Measuring the angular momentum deficit (AMD) in our systems allows for such a metric. Following \citet{Laskar1997}, the ratio of a planet's relative AMD to that of its system as a whole can be interpreted as a secular measure of the perturbation of the orbit. We can then find the regime in which the possibility of strong encounters with other bodies in the planetary system are forbidden, implying long-term stability.

Due to its conservative nature, many different stability criteria can be included into the AMD framework. For our purposes, we specifically include the criteria imposed on the critical AMD (the value at which a given planet becomes unstable) stemming from Hill stability \citep{Petit2018}, orbit-crossing collisions \citep{Laskar2017}, and first-order mean motion resonance (MMR) overlap \citep{Petit2017}. We present a complete summary of the AMD framework in Appendix \ref{subsec:Append_AMD}. 

In order to allow for an efficient and complete classification of our
planetary systems, we adopt the following prescription: if the $j$-th planet's AMD-stability coefficient \textbf{ $\beta_{j}=\mathcal{C}_{j}/C_{c}^{H}<1$}, we do not calculate $C_{c}$; if the reverse is true, then we set $\beta_{j}=\mathcal{C}_{j}/C_{c}$. 
As noted in Section 4 of \citet{Petit2018}, this ensures the system's long-term stability or lack thereof is correctly validated. Using $\beta_j$ as a post-processing indicator, we are able to correctly categorize our systems as stable (where all planets have $\beta_j<1$), unstable (where two or more planets have $\beta_j > 1$), or meta-stable (where the innermost planet is unstable but all other planets are stable).

\section{\textsc{Tycho}'s Initial Conditions}

\textsc{Tycho} aims to provide a physically accurate depiction to
the evolution of planetary systems embedded within stellar clusters. In the following subsections, we describe our specific initial conditions for this project. However, we note that these can be readily modified to suit other needs; \textsc{Tycho} is at its core built to be a versatile tool for modeling the effects of stellar cluster dynamics on multiple scales.
To ensure that our simulations can be compared with statistics drawn from
observational surveys, our initial conditions are drawn from empirical
data while allowing some room for well-constrained, fixed parameters. 

\begin{figure}[t!]
\centering{}\includegraphics[width=0.5\textwidth]{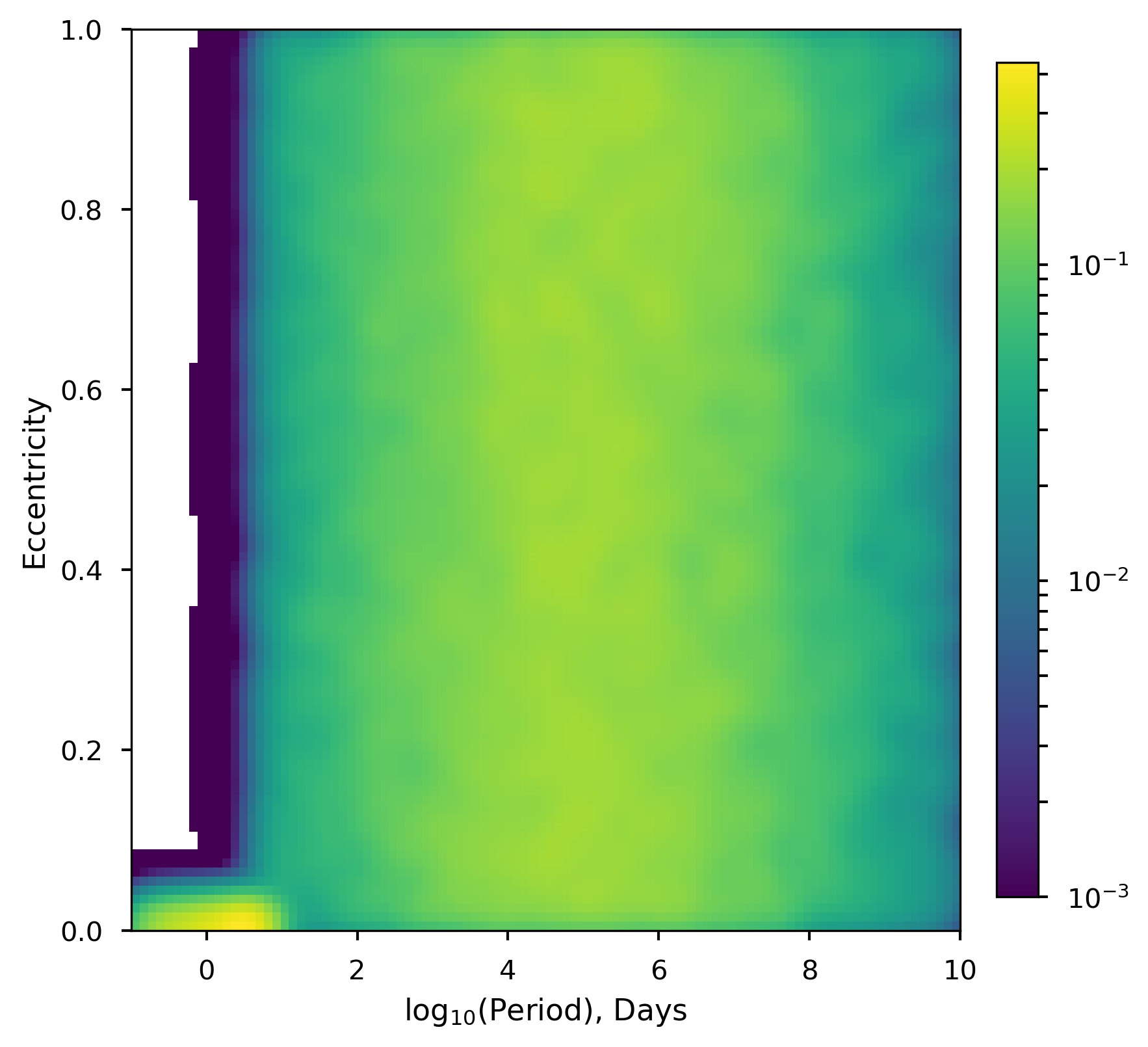}\includegraphics[width=0.5\textwidth]{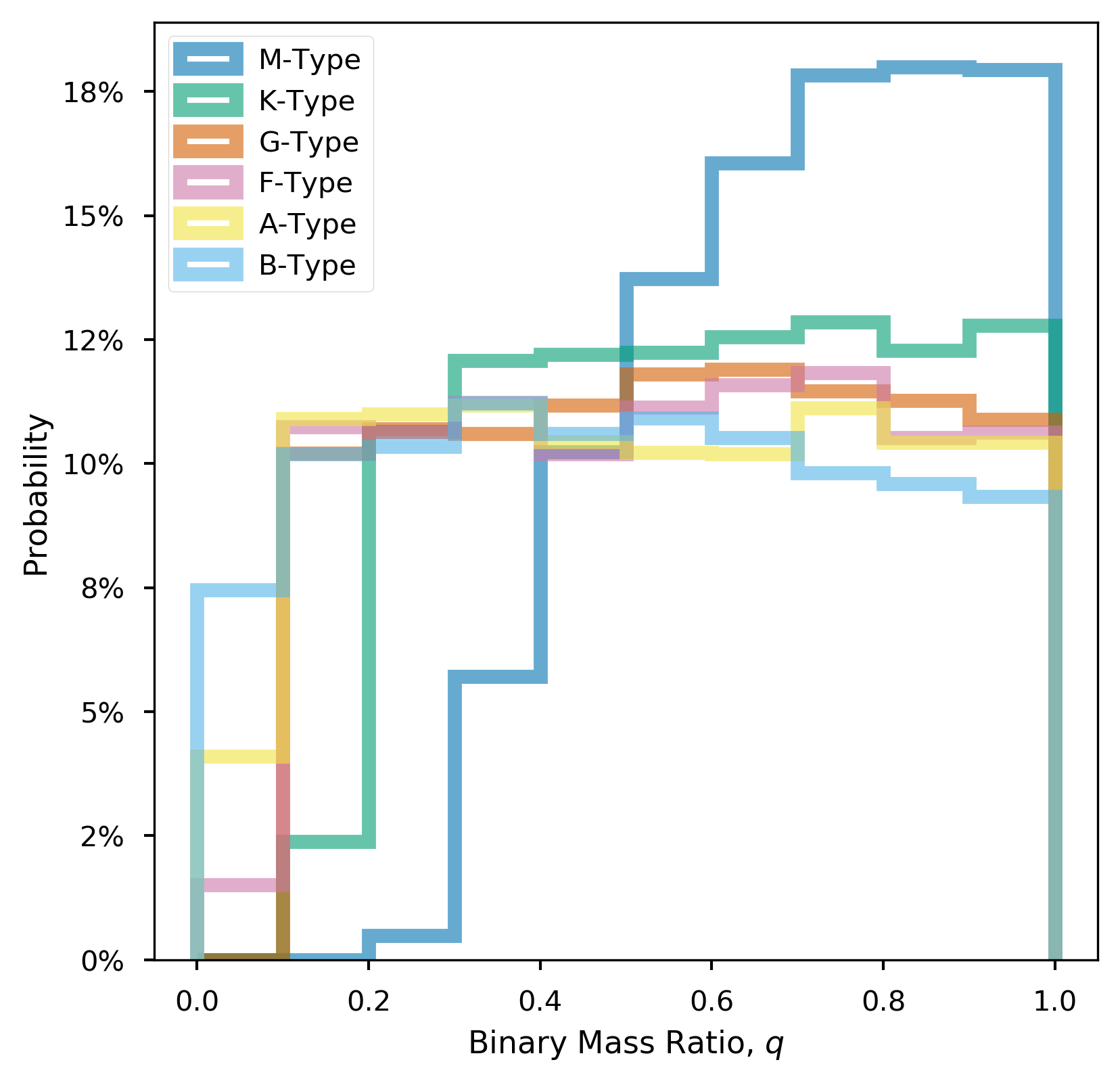}\caption{\label{fig:BinaryPop}Initial conditions for
our primordial binary population, for
a sample of 100,000 binaries. (a) The graph on the left shows
our reconstruction of the orbital distributions presented in 
\citet{Raghavan2010}
and \citep{Marks2011} for young clusters. Note, that orbits shorter
than $\sim10$ days are assumed to be circularized due to tidal effects.
(b) At right we reconstruct the binary
mass ratio, $q$. Note that the lower probability for small $q$
in M-Type stars is due to our minimum partner mass criterion, as we
treat stars separately from brown-dwarfs/planets.}
\end{figure}

\subsection{Stars and Primordial Binaries\label{subsec:Stars-and-Primordial}}

We draw our clusters' spatial density
distributions from King \citeyearpar{King1962,King1965,King1966a}
models, which provide rudimentary, if imperfect, fits to observed
stellar clusters with masses of a few times $10^{3}M_{\odot}$
\citep{ZwartMcMillanGieles2010}. Detailed observations of star-forming regions
show stellar distributions which tend to be clumpier in composition
than King models \citep{Sanchez2009,Allison2010}, with
stars still embedded in their natal gas for the first few megayears.
$N$-body models of hierarchical or fractal stellar clusters relax to
King-like spatial distributions
in a short time \citep{Smith2013,Geller2013}. We synthetically
age our stellar population and begin our models at 10 Myr, to avoid
this gas-embedded, spatially clumpy phase. In the future, we will
couple our models to detailed star-formation simulations
\citep[such as those presented in][]{Wall2019Formation}
to better capture this initial phase. For convenience, we
initialize all of our clusters on circular orbits of Galactocentric radii
9 kpc.\footnote{It should be noted that if an eccentric Galactic orbit
were to be chosen for our clusters, the spatial distribution of a cluster
described in \citet{Kupper2010} would be more appropriate.}

For our stellar masses, we sample a truncated Kroupa initial mass
function \citep[IMF;][]{Kroupa2001}. We set our lower and upper stellar
mass limits as the lowest mass for an observed red dwarf ($M_{min}=0.1M_{\odot}$; \citealp{Dieterich2014TheLimit})
and the approximate maximum single-component mass following complete
gas ejection at $\sim$5-10 Myr ($M_{max}=10M_{\odot}$; \citealp{Lada2010}),
respectively. Limiting our lower mass allows for us to more efficiently
spend computational resources integrating the cluster's evolution, without
significantly altering the overall evolution of the cluster \citep{Kouwenhoven2014}.

We draw the orbital parameters of our primordial stellar binary population
from empirical distributions derived from observational surveys.
The creation procedure implemented is as follows: 
\begin{enumerate}
\item To determine if a given primary star should have a stellar binary companion,
we adopt the empirical binary fractions as a function of primary
mass from Figure 12 of \citet{Raghavan2010}. If a star is selected to
have a secondary component, its position
becomes the center of mass of the new binary. 
\item The secondary mass is selected from the truncated Kroupa IMF and added
to the initial mass of the center of mass particle. This allows
us to recover the observed IMF when we redistribute the
mass between the primary and secondary star. 
\item Using a uniform mass-ratio distribution \citep{Raghavan2010,Goodwin2013},
we split the the center of mass particle's new total mass into the
primary and secondary star. 
\item To determine the stellar binary's orbital elements, we use the
primary's mass to draw from empirically-based distributions.
The period is generated from the log-normal distribution
of \citet{Raghavan2010} for lower-mass and \citet{Sana2012} for
higher-mass systems. Eccentricities are drawn from a uniform distribution
well documented in literature \citep{Raghavan2010,Duchene2013,Moe2017}.
\item The orientation of the system with respect to the cluster's coordinate
system is chosen randomly and uniformly over the unit sphere as
described in \citet{Arvo1992}. 
\end{enumerate}
Figure \ref{fig:BinaryPop} provides an overview of the yield
of the above described procedure. Once generated, binaries are
added to the $N$-body particle set to be picked up by \textsc{Multiples}
before the cluster is scaled to virial equilibrium. All internal
orbital interactions, close encounters and stellar evolution are tracked
and stored by \textsc{Tycho}'s bookkeeping procedures for future analysis.
There are no initial systems consisting of more than two bodies in
our simulated clusters.

\floattable
\begin{deluxetable}{lcccc}
    \tabletypesize{\footnotesize{}}
    \tablecaption{\label{tab:planets}Implemented Planetary Systems}
    \tablehead{
    \colhead{} & \colhead{Mass ($M_{J}$)} & \colhead{Eccentricity} & \colhead{Semi-Major Axis (AU)} & \colhead{Avg. AMD Stability Coefficient, $\beta_{AMD}$}
    }
    \startdata
    Terrestrial  & 0.003  & 0.016  & $1.000\times\left(\frac{M_{host}}{M_{\odot}}\right)^{2}$  & $0.9015\pm4.50\times10^{-4}$ \\
    Jovian  & 1  & 0.048  & $5.454\times\left(\frac{M_{host}}{M_{\odot}}\right)^{2}$  & $0.2394\pm2.35\times10^{-4}$ \\
    Neptunian  & 0.054  & 0.009  & $30.110\times\left(\frac{M_{host}}{M_{\odot}}\right)^{2}$  & $0.02361\pm1.79\times10^{-5}$ \\
    \enddata
\end{deluxetable}

\subsection{Planetary Systems \label{subsec:PlanetaryIC}}

Within the initial cluster models presented in the previous section,
we model the planetary systems after our own solar system---using a stable,
well-studied system allows us to differentiate the effects of star-star
scattering on multi-planet systems from internal interplanetary dynamics.
Thus we adopt a scaled version of the solar
system, which contains only analogs for Earth, Jupiter and Neptune. We refer to as Terrestrial, Jovian, and Neptunian, respectively.
We initialize our systems as coplanar, consistent with leading formation models \citep{Laskar2000}.

In accordance with current
formation theory, the placement of each system's
Jovian is scaled to the snow line of its host star \citep{Kennedy2008}.
Thus, the Jovian planets have semi-major axes
$a_{jovian}=a_{J}\left(\frac{M_{host}}{M_{\odot}}\right)^{2}$ \citep{Ida2005}.  After scaling the Jovian's semi-major axis, we
scale the other planetary orbits so
that the system remains long-term stable. By fixing the ratio of
periods between each planet and the system's Jovian, we
ensure that the solar system's innate stability
is preserved regardless of the mass of the host star. A good
measure of this is the angular momentum deficit of the planets (see Section \ref{subsec:AMD}). Table \ref{tab:planets}
summarizes our model's parameters and provides the AMD stability coefficient of each planet, averaged over the
host star masses in our simulations.

\section{Proof-of-Concept Simulations }

\subsection{Experimental Setup}

To illustrate our improved approach for simulating planetary systems
within clusters, we simulate 40 independently-seeded
open clusters with initial conditions as described above.
These consist of 10 realizations for each of the possible parameter combinations of the
number of center-of-mass objects ($N\in\{100,1000\}$) and the depth
of the King potential ($W_{0}\in\{3.0,6.0\}$).\footnote{We chose to vary only these parameters to demonstrate key
attributes of this approach. In subsequent papers we will
include initial conditions that are more directly associated with
observed star clusters.} Each realization is simulated according to our prescription
for approximately 2 Gyr, which is past the point of cluster dissolution in most cases. Planetary systems are generated only around single stars and include only the Jovians in our full-cluster $N$-body simulations
(as discussed above), but we retain the full planetary systems
in our scattering experiments. For the purposes of showcasing the dynamics, 
we limit our close encounter database to those with 
$r_{periapsis}<2 a_{outer}$.

Our simulations were  performed on the \textsc{Draco} super-computing cluster at Drexel University.  \textsc{Draco} consists of 24 compute nodes,  each consisting of 2 Intel Xeon x5650 CPU 6-core chips and 4--6 vintage (Tesla/Titan) NVIDIA GPUs,  allowing us to take advantage of \textsc{PH4}'s parallel and \textsc{cuda}-enabled design. Our $N$-body simulations took 553 CPU hours to complete, and our scattering experiments took 1690 CPU hours, for a total of ~1.33 weeks distributed across 10 nodes.

\begin{figure}[t!]
\centering{}\includegraphics[width=0.95\textwidth]{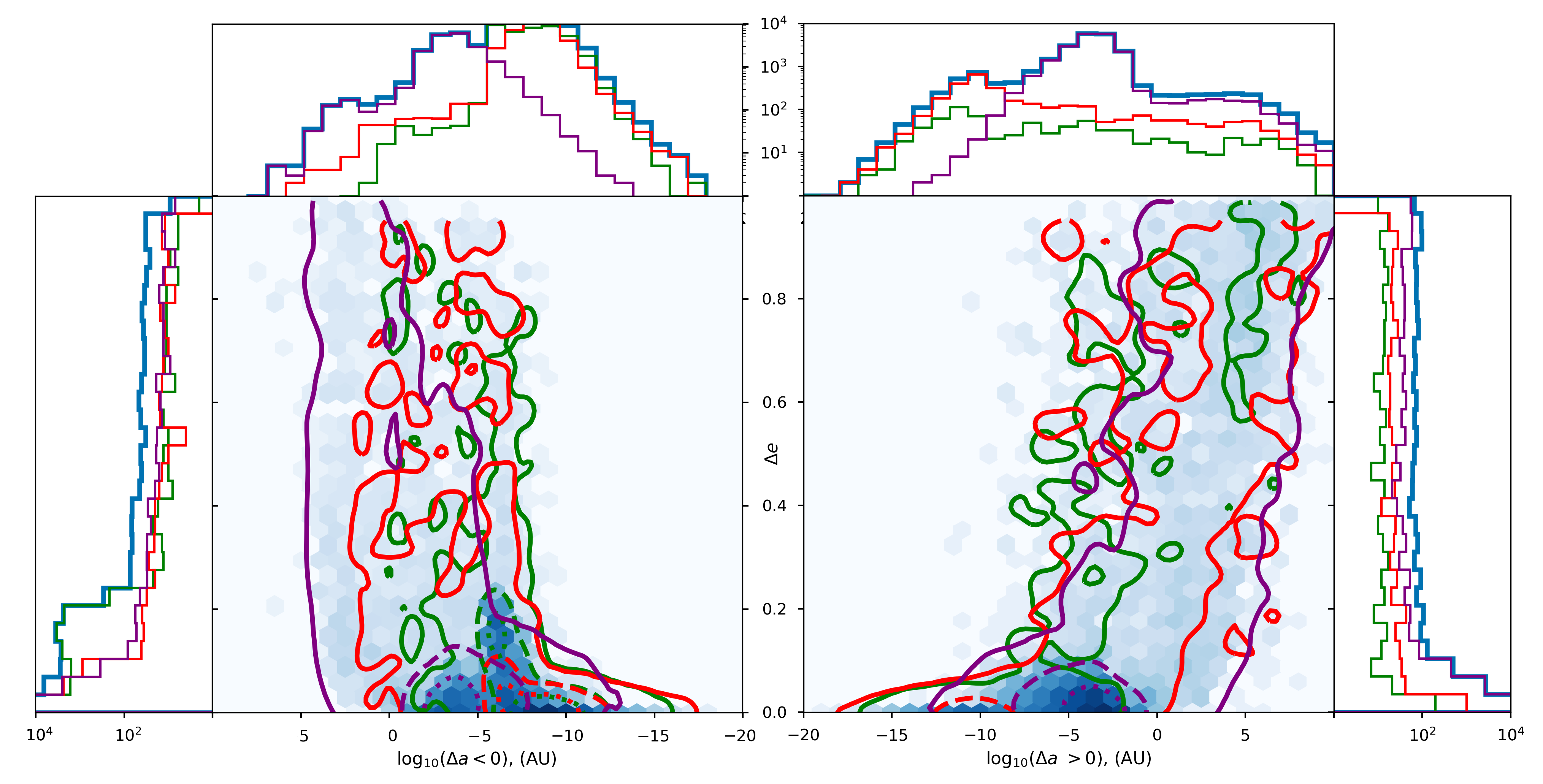}\caption{\label{fig:DeltaAvsDeltaE}Total change in
orbital elements $a$ and $e$ of exoplanets at the point of their
host star's ejection from the cluster. The background hexagon-bined
histogram is the total for all planets and the contours represent
the three planetary populations: green for Terrestrials; red for Jovians;
and purple for Neptunians. Dotted lines represent the contour containing greater than 100
counts; dashed lines represent contour containing greater than 10 counts; and solid lines represent the contour containing greater than 1 count.}
\end{figure}

\subsection{Results}

A catalog of our simulations has been provided to the publisher, containing a total of 41,313 planetary systems of various configurations. Figure \ref{fig:DeltaAvsDeltaE} shows the relation between the total change in eccentricity and the total change in semi-major axis for our three planetary populations after cluster dissolution. 
Most of our stellar population only encountered another stellar system once over the 2 Gyr evolution.  Given the much greater scattering cross section of our Neptunian population, our results confirm the expectation that they are far more significantly affected by stellar flybys than the Jovian or Terrestrial planets. Indeed, 
of the 371 planets that escaped their host system across all encounters, 222 were Neptunians, 77 were Jovians, and 61 were Terrestrials. Further, of these planets that left their systems, the Neptunians were more likely to be captured (26 Neptunians versus 5 Jovians and 0 Terrestrials).

\begin{figure}[t!]
\centering{}\includegraphics[width=0.8\textwidth]{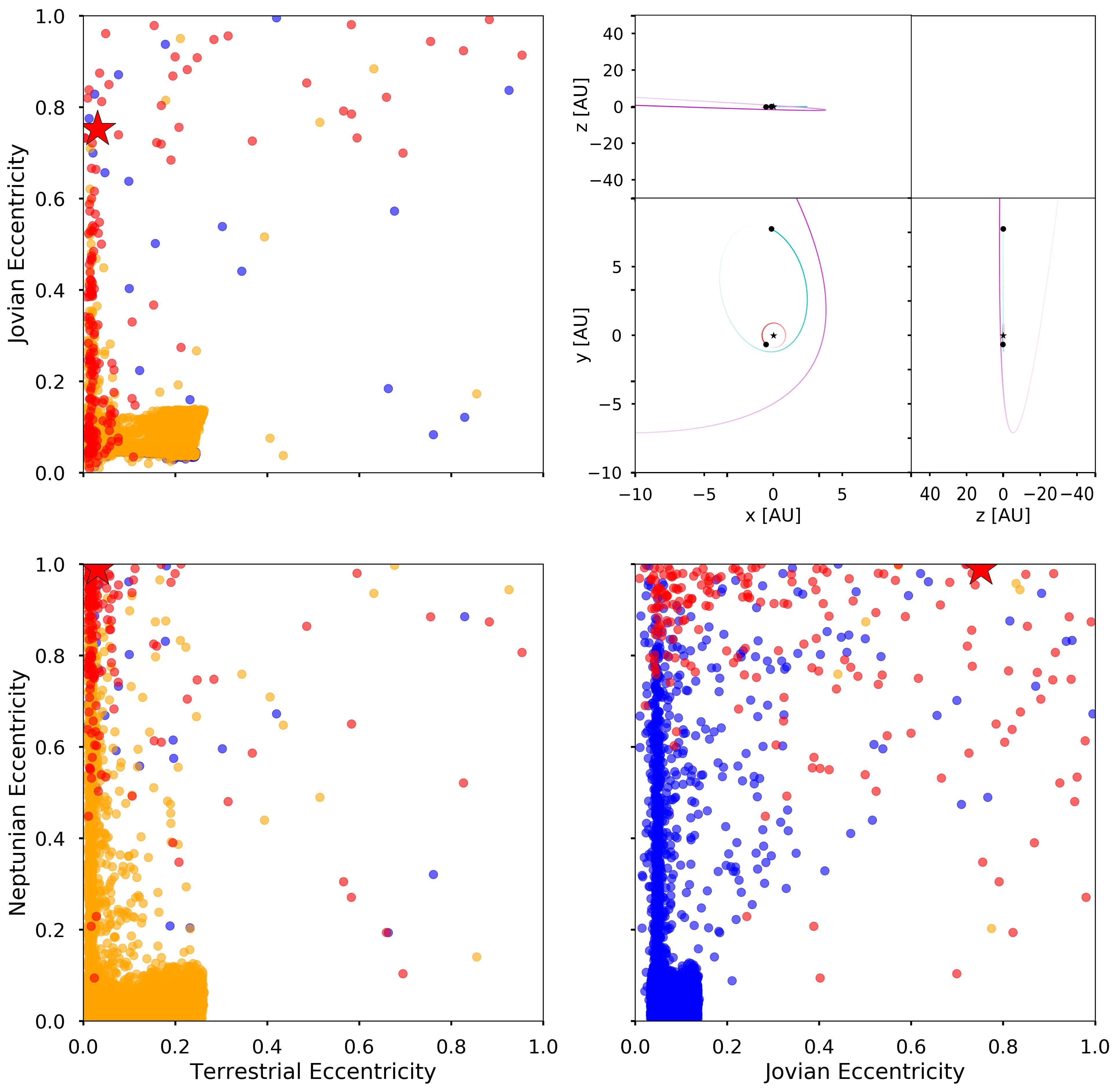}\caption{\label{fig:DeltaEvsDeltaE} The eccentricity distribution of planet pairs for systems which retained their initial structure after the encounter is finished. Systems that are long-term stable are in blue, weakly unstable in orange, and completely unstable in red. We note that there are no systems that are unstable in-which there is not a significant change to the Jovian or Neptunian orbital eccentricity. Additionally, we have included an example unstable system which is marked by a star in the eccentricity graphs. Here, the orbits are denoted as follows: Terrestrial is red; the Jovian is aqua; and the Neptunian is purple.}
\end{figure}
\begin{figure}[!p]
\centering{}\includegraphics[height=0.35\textheight]{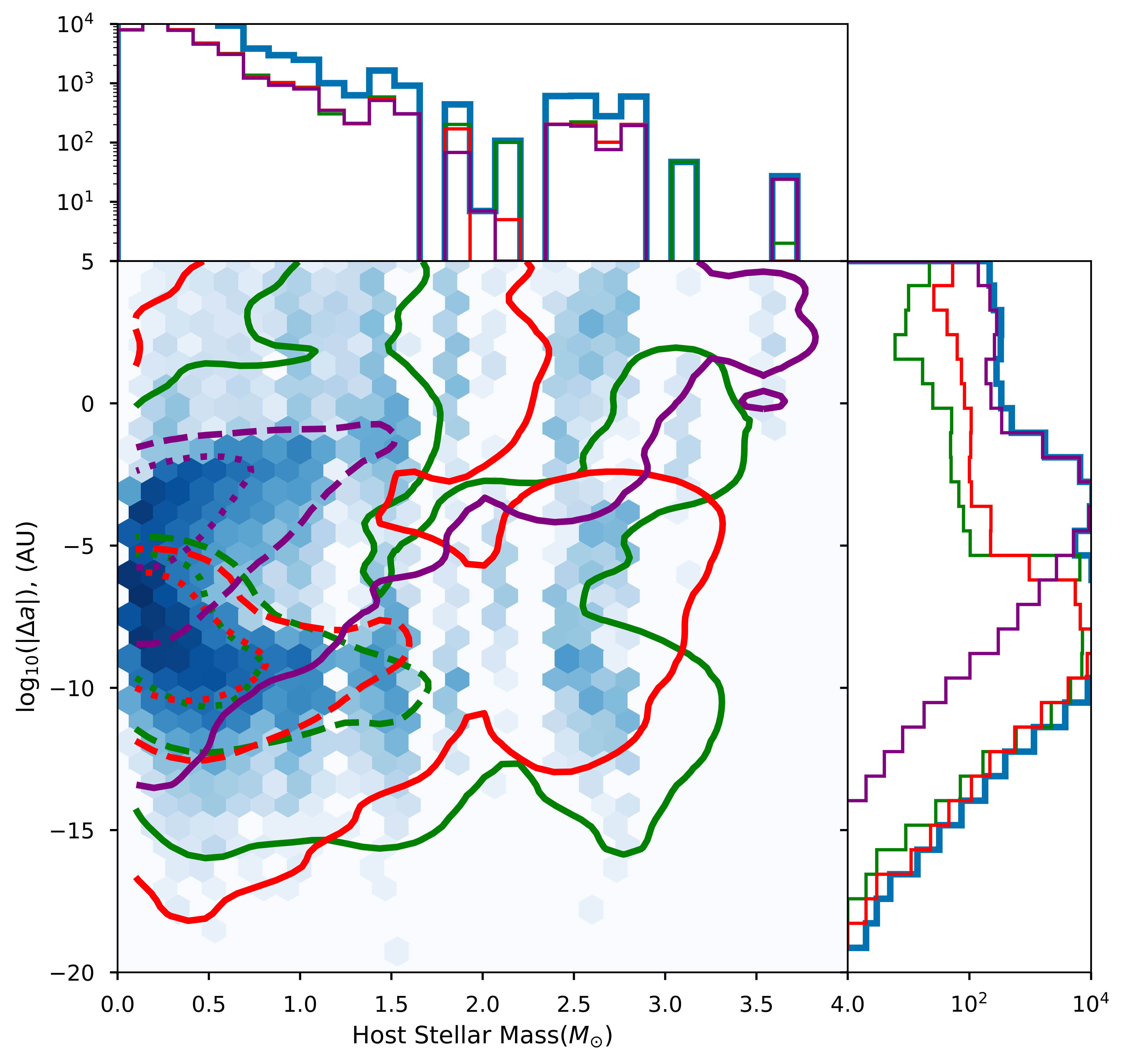}~~~~~~~\includegraphics[height=0.35\textheight]{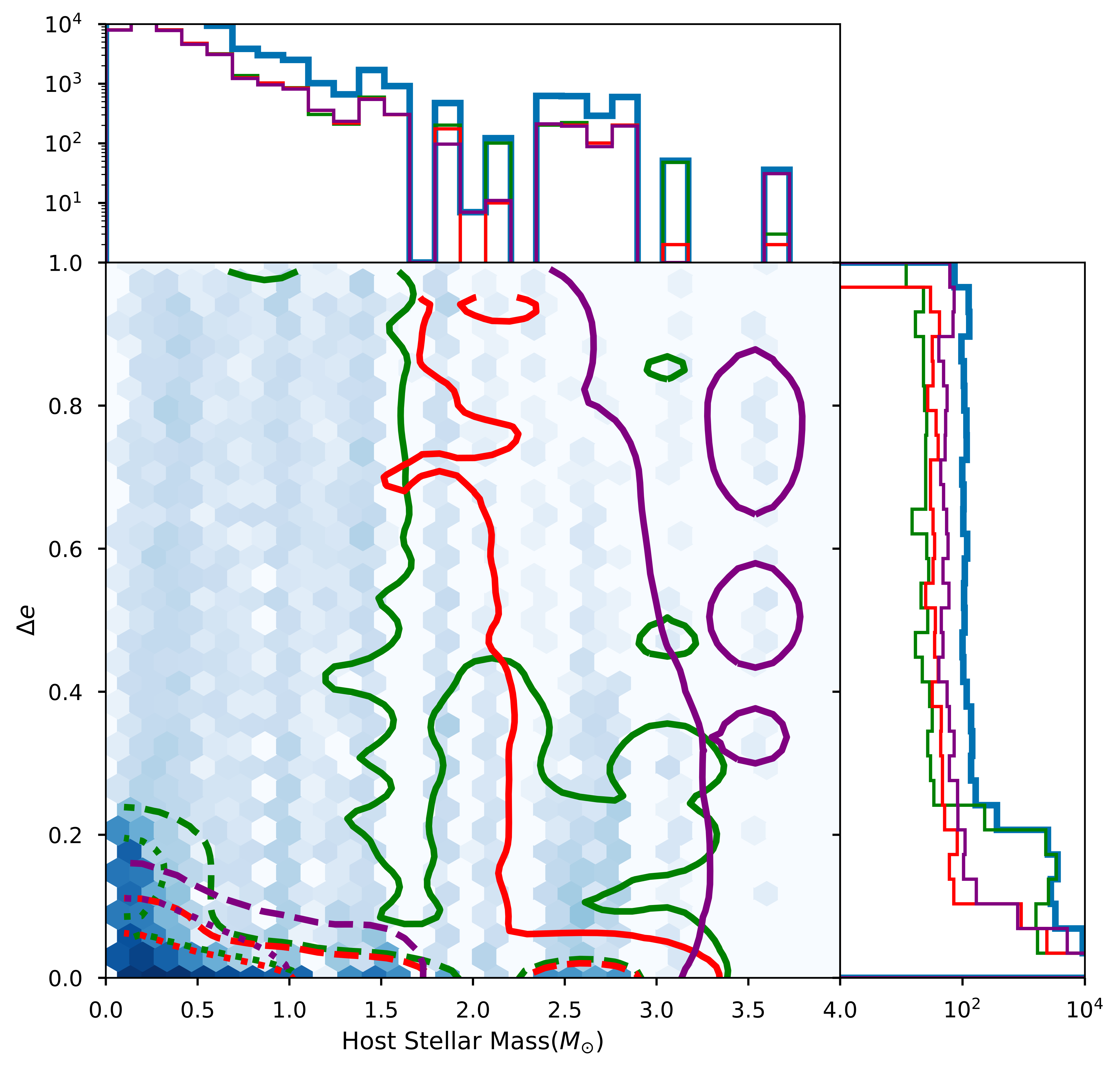}\caption{\label{fig:DeltaAvsStellarMass}Total change in semi-major axis, $a$, and eccentricity, $e$, of exoplanets at the point of their host star's ejection from the cluster as a function of stellar mass of the host
star. The background hexagon-binned histogram is the total for all
planets and the contours represent the three planetary populations:
green for Terrestrials; red for Jovians; and purple for Neptunians.
Dotted lines represent the contour containing greater than 100
counts; dashed lines represent contour containing greater than 10 counts; and solid lines represent the contour containing greater than 1 count.
}
\end{figure}
\begin{figure}
\centering{}\includegraphics[height=0.40\textheight]{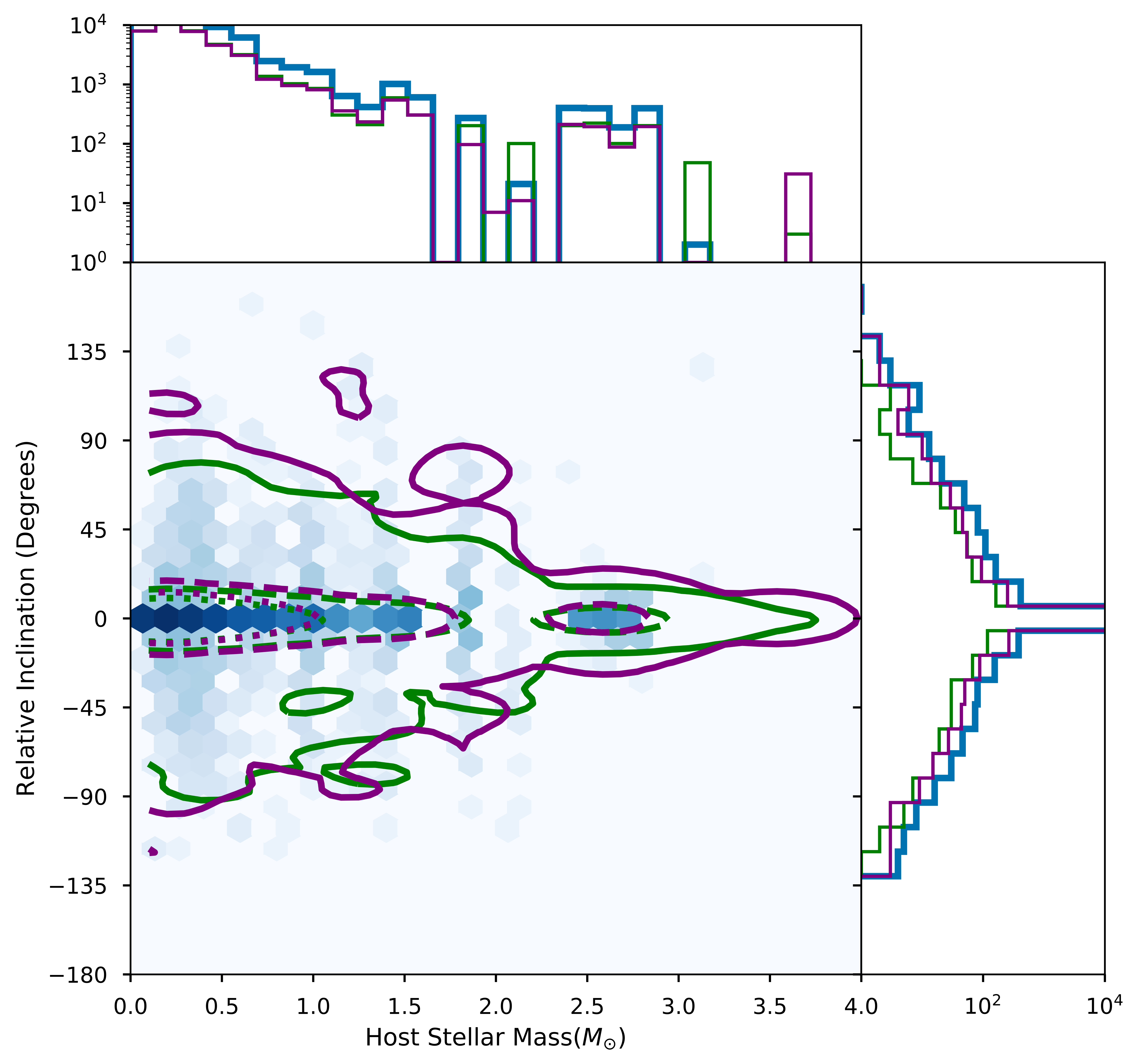}\caption{\label{fig:RelInc}Distribution of orbital inclination relative to the corresponding system's Jovian. The background hexagon-binned histogram is the total for all
planets (excluding the reference Jovians) and the contours represent the two planetary populations:
green for Terrestrials; and purple for Neptunians.
Dotted lines represent the contour containing greater than 100
counts; dashed lines represent contour containing greater than 10 counts; and solid lines represent the contour containing greater than 1 count.}
\end{figure}
\begin{figure}[t!]
\centering{}\includegraphics[width=0.5\textwidth]{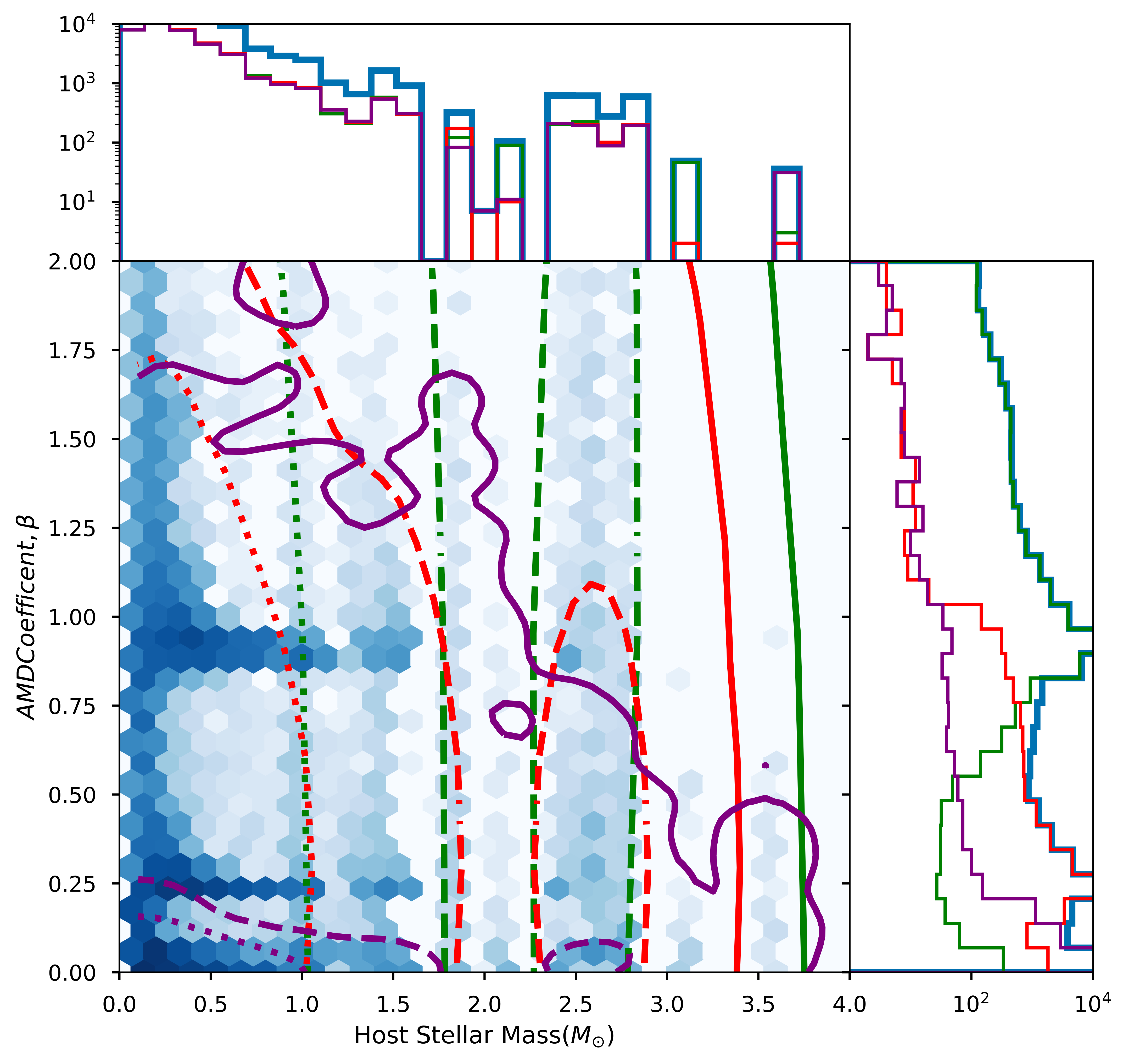}\caption{\label{fig:AMDCurve}Distribution of AMD stability coefficient
$\beta_{AMD}$ for all exoplanets at the point of their host star's
ejection from the cluster as a function of the final mass of the host
star. The background hexagon-binned histogram is the total for all
planets and the contours represent the three planetary populations:
green for Terrestrials; red for Jovians; and purple for Neptunians.
Dotted lines represent the contour containing greater than 100
counts; dashed lines represent contour containing greater than 10 counts; and solid lines represent the contour containing greater than 1 count.}
\end{figure}

We note that most of our systems do not have large changes in eccentricity or semi-major axis even though all of the systems included in our catalog have undergone at least one stellar encounter. Those systems that do experience significant changes in their semi-major axes usually experience changes in their eccentricities as well. 
We see that more of our systems that retain three planets after the encounter (as plotted in Figures~\ref{fig:DeltaAvsDeltaE}~-~\ref{fig:AMDCurve}) experience hardening (i.e., shrinking of the semi-major axis) due to stellar encounters. Apparently, encounters that soften the orbits are either less common, or are more likely to lead to the disruption of the planetary system. 

Furthermore, the eccentricity histogram of the left graph of Figure \ref{fig:DeltaAvsDeltaE} 
shows that as the system shrinks, our terrestrial population experiences large changes in eccentricity while staying near the original semi-major axis.

Figure \ref{fig:DeltaEvsDeltaE} compares the change in eccentricity in each of our planetary types drawn from systems that retain all three original planets. The color indicates the long-term stability of the system as determined by the AMD framework. We see that while all of our systems begin as AMD stable, 37.5\% become weakly unstable and 0.81\% become fully unstable after 2 Gyr. Of this unstable population, a significant portion of outer planets meet the criteria for eventual collision with an inner body rather than instability generated via the MMR mechanism, with no system becoming Hill unstable. As expected,
there is a high concentration of unstable systems in which Jupiter and Neptune both experience large changes in eccentricity. However, 
very few Jovians reach a point where their AMD stability coefficients exceed unity. In weakly stable systems, 
most arise from instability in their companion Terrestrial that results in orbital decay and inspiral into the parent star. 

To provide an example from the catalog, we have included a very unstable system in Figure \ref{fig:DeltaEvsDeltaE}. This system is the result of a stellar binary ($M_{b1} = 0.133~\mathrm{M_\odot}$ \& $M_{b2} = 0.369~\mathrm{M_\odot}$; Period $= 9746$ Days, $a= 7.105$ AU, and $e = 0.248$) encountering the host star ($M_\star = 0.950 ~\mathrm{M_\odot}$). The encounter has a closest approach of 10.001 AU over its 35.6 Kyr lifetime ($a = 1224.27$ AU and $e = 1.008$), placing the binary well within the planetary system. A summary of the final orbital elements of the system can be found in Table \ref{tab:ExampleEnc}. As a result of this encounter, the planetary system is predicted to result in an eventual collision between the Jovian and Neptunian while the Terrestrial inspirals towards the host star.

\floattable
\begin{deluxetable}{lcccc}
    \tabletypesize{\footnotesize{}}
    \tablecaption{\label{tab:ExampleEnc}Resulting System Around Star \textnumero~339}
    \tablehead{
    \colhead{} & \colhead{Eccentricity} & \colhead{Semi-Major Axis (AU)} & \colhead{Relative Inclination}& \colhead{$\beta_{AMD}$}
    }
    \startdata
    Terrestrial  & 0.031 (+0.021)  & 0.898 (+0.005)  & -18.137\textdegree & 570.68\\
    Jovian & 0.752 (+0.704)  & 4.687 (-0.232)  & -- & 3.498 \\
    Neptunian  & 0.99 (+0.981)  & 363.889 (+336.726) & +23.570\textdegree & 1.878\\
    \enddata
    \tablecomments{This system can be found in the catalog provided to the publisher under the System Key: \textsc{Adam\_N1000\_W6/S1340/Enc-0\_Rot-63}. The table includes the final value of each parameter and the difference from the initial condition, indicated in parenthesis. We take our inclination measurements from the orbital plane of the most massive planet, which in these proof-of-concept simulations is the system's Jovian.
    }
\end{deluxetable}

Finally, we turn to the relationship between changes in the orbits and the mass of the system's stellar host. 
As the stellar mass increases 
so does the change 
in the semi-major axis, with the Neptunian population strongly following this trend.
(as seen in Figure \ref{fig:DeltaAvsStellarMass}) Meanwhile, planets experience more dispersion of their eccentricity at lower stellar host masses. As expected, the relative inclinations of the planetary systems resemble that of a Gaussian distribution centered about the co-planar starting condition~($\sigma_T=4.7$\textdegree ~\& $\sigma_N = 5.9 $\textdegree; see Figure \ref{fig:RelInc}). 
Additionally, Figure \ref{fig:AMDCurve} shows that planets orbiting lower mass stars have the highest amount of dispersion in their AMD stability coefficient. Note medians of the three planetary types stay nearly constant across the range of host masses (centered around $\beta_T$ = 0.90, $\beta_J=0.24$, and $\beta_N=0.02$).

\section{Discussion and Closing Remarks}

\subsection{Discussion}
The results from these simulations are revealing; they predict that a significant portion of planets within gas-ejected open clusters experience long-term stability changes due to stellar close encounters. From planetary ejections to inner-planet inspiral, it is clear that encounters fitting the parameters described in Sections 2.4 and 4.1 can drive dynamical changes in the structure of the planetary system.

As previously noted, we see that systems that become unstable always have a Neptune with a significant eccentricity change. This is expected, as when stellar systems encounter one another, the outer planets will have the largest orbital perturbation. By increasing the orbital eccentricities, these changes lower the outer planets' periapses, resulting in further orbital shifts via planet-planet interactions and, eventually, long-term instability.

We can understand the trends in our planetary systems' orbital changes through the lens of dynamical mass segregation \citet[]{McMillan2007AClusters}. As an open cluster evolves, stellar encounters
redistribute energy throughout the cluster, attempting to 
drive the stars toward energy equipartition.  This leads the more massive stars to move more slowly and hence sink toward the center of the cluster.  As a result, more massive stars are more likely to encounter other massive stars, leading to larger encounter energies and thus greater perturbations to a planet's semi-major axis. Additionally, lower-mass stars pass through this dense core before being ejected from the cluster via evaporation, often times after a single strong encounter. During such strong encounters (which our cuts highlight), even tightly-packed planetary systems around these low-mass hosts can experience dramatic orbital perturbations.

Regarding the significant population of unstable Terrestrials, we note that these planets are near the critical point of $\beta=1$ before any encounters. As such, only a small change in the angular momentum of the system is required to push this category of planet towards an instability criterion either via inspiral into its stellar host or a collision in densely packed system. We find that while the median of our Terrestrial population remains at $\beta=0.945$, they exceed $\beta=1$ at the 61.63 percentile. Conversely, the Jovians and Neptunian population only exceed $\beta=1$ at the 99.5 percentile despite their tendency towards significant orbital change.

\subsection{Future Work}
In future papers we will improve and expand
our initial conditions to better match observed clusters, such as
those presented in Table \ref{tab:cluster_cat}, and implement new
numerical tools to enhance \textsc{Tycho}'s capabilities in order to make more quantitative predictions about the expected observational
sample of exoplanets.

As mentioned in Section \ref{subsec:Modeling-Dynamics}, our current
methodology does not actively simulate planetary systems between stellar
encounters. While this does not affect the validity
of our results thanks to the robustness of the AMD stability
criterion, integrating a system over long time scales will expose important
nuances in its evolution. Recently, a generalized
hierarchical approach to the long-term secular integration of planetary
systems was described by \citet{Hamers2016}. It was further expanded
to include a secular treatment of close encounters between such hierarchical
systems and is now included in AMUSE (see \textsc{SecularMultiples}
community code; \citealp{Hamers2018}). To improve both \textsc{Tycho}'s
physical accuracy and computational efficiency, we plan to integrate this
approach into our scattering experiments in the next iteration.

Additionally, Tycho does not follow planetary systems through multiple stellar encounters. Rather, it treats encounters as wholly separate from one another, regardless of a system's history or future. 
Most stars in these clusters only encounter one other star during the cluster lifetime and thus Tycho's approach is appropriate for the simulations presented here. However, in future larger simulations, the number of stars that undergo several encounters in their lifetime, and particularly those that migrate to the center of the cluster during the core collapse epoch, will not be insignificant.
The implementation of \textsc{SecularMultiples} will allow planetary systems that undergo multiple encounters to be efficiently
evolved between encounters, and therefore enable proper handling of such encounter chains.

\subsection{Conclusions}
In this paper, we 
present a robust methodology, \textsc{Tycho}, that aims to solve the multifaceted problem of evolving planetary systems from their birth in their natal clusters to their eventual dispersion into the Galactic disk. Our proof-of-concept simulations verify that \textsc{Tycho} can simulate such clusters within a reasonable amount of time without compromising the physical integrity required for such work.

Examining the results of our proof-of-concept runs, we find that a significant portion ($37.5\%$) of our remaining solar-system analogs become weakly unstable due to close encounters with other stellar systems. A smaller portion become fully unstable, mainly due to the collisional criterion with the inner planet. Additionally, 
while systems orbiting high-mass stars experience larger changes in their semi-major axes than those around lower-mass stars, systems around low-mass stars experience larger dispersion in $\beta_{j}$. Finally, 
0.8\% of the planets in our population were ejected from their host system and/or captured by another stellar system due to stellar close encounters.

In summary, we have developed a methodology that allows us to explore how stellar birth environments influence the evolution of planetary systems. Building upon previous numerical solutions, \textsc{Tycho} strikes a balance between computational efficiency of scattering experiments and direct $N$-body simulations. Furthermore, we plan to continue developing \textsc{Tycho} to enable direct comparisons between simulated and observational planetary populations, the progress of which can be followed on the project's GitHub repository (\url{https://github.com/JPGlaser/Tycho}).


\acknowledgements{This research has been supported by Drexel University's College of Arts and Science under their Doctoral Research Fellowship and their STAR program. This research made use of the NASA Exoplanet Archive, which is operated by the California Institute of Technology, under contract with the National Aeronautics and Space Administration under the Exoplanet Exploration Program. Computational resources were supported by Drexel's University Research Computing Facility through NSF Award AST-0959884. A. Geller acknowledges support from NSF Astronomy and Astrophysics Postdoctoral Fellowship Award AST-1302765. The authors would also like to acknowledge Adam Dempsey (CIERA, Northwestern University) for his help in discussing the AMD stability criteria, and the AMUSE User Group.}

\appendix

\section{Summary of the Relative Angular Moment Deficit (AMD) Framework }
\label{subsec:Append_AMD}
We summarize here some details of the literature on the AMD framework formalism.
We assume a planetary system of $N_{p}$ planets orbiting a single star of mass $M_{s}$, where we may work within the heliocentric reference frame such that each planet of mass, $m_n$ has an orbit that can be described by $(a_n, e_n, i_n, \nu_n, \omega_n, \Omega_n)$ and that the total linear momentum is zero.\footnote{Allowing the following definitions: $a_n$ is the semi-major axis, $e_n$ the eccentricity, $i_k$ the inclination, $\nu_n$ the true anomaly, $\omega_n$ is the argument of periapsis, and $\Omega_n$ is the longitude of the ascending node.} Further, we assume that the normal to the reference plane is in the $z$ direction. Thus, the vector norm of the total angular momentum is:
\[G = \sum^{N_p}_{n=0}{||\textbf{r}_n\wedge \dot{\textbf{r}}_n||} = 
\sum^{N_p}_{n=0}{\Lambda_n \sqrt{1-e_n^2}cos(i_n)}\],
where $\Lambda_n = m_n \sqrt{G M_s a_n}$.  The AMD of the system, $C$, is defined as ``the difference between the difference between the norm of the angular momentum of a coplanar and circular system with the same semi-major axis values and the norm of the angular momentum (G)'' \citep{Laskar2000,Laskar2017}. Therefore, the relative AMD of the $j$-th planet in the system is defined as: 
\begin{equation}
\mathcal{C}_{j}=\frac{C}{\Lambda_{j}}=\sum_{n=1}^{N_{p}}\frac{m_{n}}{m_{j}}\sqrt{\frac{a_{n}}{a_{j}}}\left(1-\sqrt{1-e_{n}^{2}}\cos\left(i_{n}\right)\right).
\end{equation}
For each pair of planets within the system, there are a series of stability criteria which
can be imposed within the AMD framework; each resulting in a specific
critical AMD, $C_{c}$, where the system becomes long-term unstable if at any point for any planet $\mathcal{C}_{j}>C_{c}$. We therefore define the AMD stability coefficient as $\beta_{j}=\mathcal{C}_{j}/C_{c}$, where $\beta_j>1$ results in an unstable system.

First, we consider the possibility of planetary collisions and first-order effects of mean motion resonances (MMR) between planet pairs. As detailed in \citet{Laskar2017} and \citet{Petit2017}, when taking those effects into account the critical AMD can be expressed as:
\begin{equation}\label{eq:critC}
C_{c}(\alpha, e_c, e'_c)=\begin{cases}
\gamma\sqrt{\alpha}\left(1-\sqrt{1-e_{c}^{2}}\right)+\left(1-\sqrt{1-e_{c}^{\prime2}}\right) & ~~~~~(\alpha<\alpha_{R})\\\\
{\textstyle\frac12} g(\alpha,\varepsilon)^{2} \gamma\sqrt{\alpha}\,/\,(1+\gamma\sqrt{\alpha}) & ~~~~~(\alpha\geq\alpha_{R})
\end{cases}
\end{equation}
allowing:
\begin{equation}
\varepsilon=\frac{m+m^{\prime}}{M_{s}},
\quad\gamma=\frac{m}{m^{\prime}}\,
\quad\alpha=\frac{a}{a^{\prime}}=\left(\frac{P}{P^{\prime}}\right)^{2/3}.
\end{equation}
The piecewise nature of $C_c$ comes from the dual stability conditions imposed ($\alpha < \alpha_R$ relates to collisional conditions and $\alpha \geq \alpha_R$ relates to MMR conditions). It follows that there exists an $\alpha_R$ such that Equation \ref{eq:critC} is continuous. We note that it makes little sense to take into account MMR effects for $\alpha_R<0.63$ which corresponds to the 2:1 resonance. Using this boundary condition and evaluating the collision-based criteria of $C_c$ with $\alpha\approx 1$ presented in \citet{Laskar2017}, $\alpha_R$ becomes the solution to the following equation: $3^{6}\left(1-\alpha_{R}\right)^{7}-3^{2}2^{9}\left(1-\alpha_{R}\right)^{3}r\varepsilon-2^{14}\left(r\varepsilon\right)^{2}=0$.

For $\alpha<\alpha_R$, we will need to calculate the critical eccentricities which are such that the inner $(m,a,e)$ and outer $(m^{\prime},a^{\prime},e^{\prime})$ planets would collide. These can be obtained from the following set of equations:
\begin{equation}
(e_{c},e_{c}^{\prime})=\begin{cases}
\alpha e_{c}+\frac{\gamma e_{c}}{\sqrt{\alpha\left(1-e_{c}^{2}\right)+\gamma^{2}e_{c}^{2}}}-1+\alpha=0 & ~~~~~(e_{c}\in[0,1])\\\\
\alpha e_{c}+e_{c}^{\prime}-1+\alpha=0 & ~~~~~(e_{c}^{\prime}\in[0,1])
\end{cases}
\end{equation}

For the case when $\alpha \geq \alpha_R$, we must consider the instabilities caused by MMR overlaps. As presented in \citet{Petit2017}, the square of the normalized minimal AMD to enter a resonance, $\sqrt{c_{min}}$, can be written as the piecewise function, $g(\alpha,\varepsilon)$, defined as:
\begin{equation}
g(\alpha,\varepsilon)=\begin{cases}
\frac{3^{4}}{2^{9}}\frac{(1-\alpha)^{5}}{r\varepsilon}-\frac{32}{9}\frac{r\varepsilon}{(1-\alpha)^{2}} & ~~~~~(\alpha<\alpha_{circ})\\
0 & ~~~~~ (\alpha\geq\alpha_{circ})
\end{cases}
\end{equation}
where:
\begin{equation}
 r=\frac{K_{1}\left(\frac{2}{3}\right)+2K_{0}\left(\frac{2}{3}\right)}{\pi},
\quad\alpha_{circ}=1-\frac{4}{3^{6/7}}\left(r\varepsilon\right)^{2/7}
\end{equation}
with $K_n(x)$ being the modified Bessel function of the second kind. The function $g(\alpha,\varepsilon)$ goes to zero after $\alpha\geq\alpha_{circ}$  because this is the point where all orbits (including circular ones) incur MMR overlap and thus there are no configuration of stable orbits after this point.

Finally, we consider another criterion that is important to consider for closely packed planetary systems with large mass differences: Hill stability.  As presented in \citet{Petit2018}, we find that the critical AMD is
\begin{equation}
C_{c}^{H}=\gamma\sqrt{\alpha}+1-\left(1-\gamma\right)^{3/2}\sqrt{\frac{\alpha}{\gamma+\alpha}\left(1+\frac{3^{4/3}\varepsilon^{2/3}\gamma}{\left(1+\gamma\right)^{2}}\right)}+\mathcal{O}\left(\varepsilon\right).
\end{equation}
This criterion is notably stricter than our previously presented $C_c$. As such, Hill stability may be used to quickly check if a system is AMD stable. If such a system is not Hill stable, it is still possible that it can be stabilized by the collision criterion until the point at which MMR overlap dominates ($\alpha \geq \alpha_R)$ where there exists orbits that are chaotic but still satisfy the Hill stability. For visual guidance on this logic, see Figures 4 and 5 of \citet{Petit2018}.

\newpage
\clearpage{}

\end{document}